\documentclass[11pt,preprint]{aastex}
\usepackage{rotating}
 
\newcommand{\pivec}{\mbox{\boldmath $\pi$}}
\newcommand{\xivec}{\mbox{\boldmath $\xi$}}

\begin{document}
\title{OGLE-2005-BLG-018: Characterization of Full Physical and 
Orbital Parameters of a Gravitational Binary Lens}

\author{
Shin, I.-G.\altaffilmark{11},
Udalski, A.\altaffilmark{61,12},
Han, C.\altaffilmark{62,11,51},
Gould, A.\altaffilmark{62,13},
Dominik, M.\altaffilmark{64,14},
Fouqu\'e, P.\altaffilmark{63,15},
AND\\         
Kubiak, M.\altaffilmark{12},
Szyma\'nski, M.\ K.\altaffilmark{12},
Pietrzy\'nki, G.\altaffilmark{12,16},
Soszy\'nski, I.\altaffilmark{12},
Ulaczyk, K.\altaffilmark{12},
Wyrzykowski, \L.\altaffilmark{12,17}\\
(The OGLE Collaboration),\\
DePoy, D.\ L.\altaffilmark{18},
Dong, S.\altaffilmark{19},
Gaudi, B.\ S.\altaffilmark{13},
Lee, C.-U.\altaffilmark{20},
Park, B.-G.\altaffilmark{20},
Pogge, R.\ W.\altaffilmark{13}\\
(The $\mu$FUN Collaboration),\\
Albrow, M.\ D.\altaffilmark{21},
Allan, A.\altaffilmark{22},
Beaulieu, J.\ P.\altaffilmark{23},
Bennett, D.\ P.\altaffilmark{24},
Bode, M.\altaffilmark{25},
Bramich, D.\ M.\altaffilmark{26},
Brillant, S.\altaffilmark{27},
Burgadorf, M.\altaffilmark{25,28},
Caldwell, J.\ A.\ R.\altaffilmark{29},
Calitz, H.\altaffilmark{30},
Cassan, A.\altaffilmark{23},
Cook, K.\ H.\altaffilmark{31},
Corrales, E.\altaffilmark{32},
Coutures, Ch.\altaffilmark{23},
Desort, N.\altaffilmark{32},
Dieters, S.\altaffilmark{33},
Dominis Prester, D.\altaffilmark{34},
Donatowicz, J.\altaffilmark{35},
Fraser, S.\ N.\altaffilmark{36},
Greenhill, J.\altaffilmark{33},
Hill, K.\altaffilmark{33},
Hoffman, M.\altaffilmark{30},
Horne, K.\altaffilmark{37},
J\"{o}rgensen, U.\ G.\altaffilmark{38},
Kane, S.\ R.\altaffilmark{39},
Kubas, D.\altaffilmark{27},
Marquette, J.\ B.\altaffilmark{23},
Martin, R.\altaffilmark{40},
Meintjes, P.\altaffilmark{30},
Menzies, J.\altaffilmark{41},
Mottram, C.\altaffilmark{36},
Naylor, T.\altaffilmark{22},
Pollard, K.\ R.\altaffilmark{21},
Sahu, K.\ C.\altaffilmark{42},
Snodgrass, C.\altaffilmark{45},
Steele, I.\altaffilmark{25},
Vinter, C.\altaffilmark{38},
Wambsganss, J.\altaffilmark{43},
Williams, A.\altaffilmark{40},
Woller, K.\altaffilmark{44}\\
(The PLANET/RoboNet Collaborations)\\
}

\altaffiltext{11}{Department of Physics, Chungbuk National University, Cheongju 361-763, Republic of Korea}
\altaffiltext{12}{Warsaw University Observatory, Al. Ujazdowskie 4, 00-478 Warszawa, Poland}
\altaffiltext{13}{Department of Astronomy, Ohio State University, 140 W. 18th Ave., Columbus, OH 43210, USA}
\altaffiltext{14}{SUPA School of Physics and Astronomy, University of St. Andrews, KY16 9SS, UK}
\altaffiltext{15}{IRAP, Universit\'e de Toulouse, CNRS, 14 av. E. Belin, 31400 Toulouse, France}
\altaffiltext{16}{Departamento de Fisica, Universidad de Concepci\'on, Casilla 160-C, Concepci\'on, Chile}
\altaffiltext{17}{Institute of Astronomy, University of Cambridge, Madingley Road, Cambridge, CB3 0HA, UK}
\altaffiltext{18}{Department of Physics, Texas A\&M University, College Station, TX 77843-4242, USA}
\altaffiltext{19}{Institute for Advanced Study, Einstein Drive, Princeton, NJ 08540, USA}
\altaffiltext{20}{Korea Astronomy and Space Science Institute, 61-1 Hwaam-Dong, Yuseong-Gu, Daejeon 305-348, Korea}
\altaffiltext{21}{Department of Physics and Astronomy, University of Canterbury, Private Bag 4800, Christchurch 8020, New Zealand}
\altaffiltext{22}{School of Physics, University of Exeter, Stocker Road, Exeter EX4 4QL, UK}
\altaffiltext{23}{Institut d'Astrophysique de Paris, CNRS, Université Pierre et Marie Curie UMR7095, 98bis Boulevard Arago, 75014 Paris, France}
\altaffiltext{24}{Department of Physics, University of Notre Dame, Notre Dame, IN 46556, USA}
\altaffiltext{25}{Deutsches SOFIA Institut, Universitaet Stuttgart, Pfaffenwaldring 31, 70569 Stuttgart, Germany} 
\altaffiltext{26}{European Southern Observatory, Karl-Schwarzschild-Stra©¬e 2, 85748 Garching bei M\"{u}nchen, Germany}
\altaffiltext{27}{European Southern Observatory, Casilla 19001, Vitacura 19, Santiago, Chile}
\altaffiltext{28}{SOFIA Science Center/NASA Ames Research Center Mail Stop N211-3 Moffett Field CA 94035, USA}
\altaffiltext{29}{McDonald Observatory, 16120 St Hwy Spur 78 \#2, Fort Davis, TX 79734, USA}
\altaffiltext{30}{Department of Physics/Boyden Observatory, University of the Free State, Bloemfontein 9300, South Africa}
\altaffiltext{31}{Lawrence Livermore National Laboratory, IGPP, P. O. Box 808, Livermore, CA 94551, USA}
\altaffiltext{32}{Institut d\'Astrophysique de Paris, CNRS, Universit\'e Pierre et Marie Curie UMR7095, 98bis Boulevard Arago, 75014 Paris, France}
\altaffiltext{33}{University of Tasmania, School of Maths and Physics, Private bag 37, GPO Hobart, Tasmania 7001, Australia}
\altaffiltext{34}{Department of Physics, University of Rijeka, Omladinska 14, 51000 Rijeka, Croatia}
\altaffiltext{35}{Department of Computing, Technical University of Vienna, Wiedner Hauptstrasse 10, Vienna, Austria}
\altaffiltext{36}{Astrophysics Research Institute, Liverpool John Moores University, Liverpool CH41 1LD, UK}
\altaffiltext{37}{SUPA, University of St Andrews, School of Physics \& Astronomy, North Haugh, St. Andrews, KY16 9SS, UK}
\altaffiltext{38}{Niels Bohr Institute, Astronomical Observatory, Juliane Maries Vej 30, DK-2100 Copenhagen, Denmark}
\altaffiltext{39}{NASA Exoplanet Science Institute, Caltech, MS 100-22, 770 South Wilson Avenue, Pasadena, CA 91125, USA}
\altaffiltext{40}{Perth Observatory, Walnut Road, Bickley, Perth 6076, Australia}
\altaffiltext{41}{South African Astronomical Observatory, P. O. Box 9 Observatory 7935, South Africa}
\altaffiltext{42}{Space Telescope Science Institute, Baltimore, MD, USA}
\altaffiltext{43}{Astronomisches Rechen-Institut, Zentrum f\"{o}r Astronomie, Heidelberg University, M\"{o}nchhofstr. 12-14, 69120 Heidelberg, Germany}
\altaffiltext{44}{Niels Bohr Institute, University of Copenhagen, Juliane Maries Vej 30, 2100 Copenhagen, Denmark }
\altaffiltext{45}{Max Planck Institute for Solar System Research, Max-Planck-Str. 2, 37191 Katlenburg-Lindau, Germany}
\altaffiltext{61}{Optical Gravitational Lens Experiment (OGLE)}
\altaffiltext{62}{Microlensing Follow Up Network ($\mu$FUN)}
\altaffiltext{63}{Probing Lensing Anomalies NETwork (PLANET)}
\altaffiltext{64}{RoboNet}
\altaffiltext{51}{Corresponding author}


\begin{abstract}
 We present the analysis result of a gravitational binary-lensing event
OGLE-2005-BLG-018. The light curve of the event is characterized by
2 adjacent strong features and a single weak feature separated from the
strong features.  The light curve exhibits noticeable deviations from 
the best-fit model based on standard binary parameters.  To explain the 
deviation, we test models including various higher-order effects of the 
motions of the observer, source, and lens. From this, we find that it 
is necessary to account for the orbital motion of the lens in describing 
the light curve. From modeling of the light curve considering the parallax 
effect and Keplerian orbital motion, we are able to measure not only the 
physical parameters but also a complete orbital solution of the lens 
system. It is found that the event was produced by a binary lens located 
in the Galactic bulge with a distance $6.7\pm 0.3$ kpc from the Earth. 
The individual lens components with masses $0.9\pm 0.3\ M_\odot$ and 
$0.5\pm 0.1\ M_\odot$ are separated with a semi-major axis of $a=2.5
\pm 1.0$ AU and orbiting each other with a period $P=3.1 \pm 1.3$ yr.  
The event demonstrates that it is possible to extract detailed information 
about binary lens systems from well-resolved lensing light curves.
\end{abstract}

\keywords{binaries: general -- gravitational lensing: micro}


\section{Introduction}

Microlensing occurs when an astronomical object approaches close to the
line of sight toward a background star. Due to the gravity of the
intervening object (lens), light rays from the background star (source)
bend, causing splits and distortions of the source star image. For
Galactic microlensing events, the separation between the split images is
an order of milli-arcsec and thus it is difficult to directly observe
the split images. However, the lensing phenomenon can be observed by the
change of the source star brightness. For a point-source single-lens event, 
the magnification of the source star flux is represented by \citep{paczynski86}
\begin{equation}
A={u^2+2\over u(u^2+4)^{1/2}};\qquad
u=\left[ \left( {t-t_0\over t_{\rm E}}\right)^2+u_0^2\right]^{1/2},
\label{eq1}
\end{equation}
where $u$ is the lens-source separation in unit of the angular Einstein 
radius $\theta_{\rm E}$, $t_0$ is the time of closest lens-source approach, 
$u_0$ is the lens-source separation at that moment, and $t_{\rm E}$ is the 
time required for the source to transit $\theta_{\rm E}$ (Einstein time 
scale). The Einstein radius is related to the physical parameters of the 
lens system by
\begin{equation}
\theta_{\rm E}=(\kappa M \pi_{\rm rel})^{1/2},
\label{eq2}
\end{equation}
where $\kappa = 4G/(c^{2} \rm AU)$, $M$ is the mass of the lens, 
$\pi_{\rm rel}={\rm AU}(D_{\rm L}^{-1}-D_{\rm S}^{-1})$, and $D_{\rm L}$
and $D_{\rm S}$ are the distances to the lens and source, respectively. 
A standard single-lens event is characterized by a non-repeating, smooth, 
and symmetric light curve and modeling it requires 3 parameters of $t_0$, 
$u_0$, and $t_{\rm E}$. Since the first detections by \citet{alcock93} 
and \citet{udalski93}, microlensing events have been detected toward 
various star fields including the Galactic bulge \citep{udalski05,sumi10}, 
Large and Small Magellanic Clouds \citep{wyrzykowski09, wyrzykowski10}, 
and M31 \citep{calchinovati10}. Currently, events are being detected 
with a rate of more than 500 events per year, mostly toward the bulge 
field.

Among lensing events, a fraction of events are produced by lenses composed
of two masses. These binary-lens events can exhibit light curves that are 
dramatically different from those of single-lens events \citep{maopaczynski91}.  
The most prominent features occur when the source closely approaches or 
crosses caustics, which represent the set of source positions at which the 
lensing magnification of a point source becomes infinity. Describing a 
standard binary-lens light curve requires to include three additional 
parameters: the mass ratio of the companion to its host, $q$, the projected 
separation between the lens components in units of the Einstein radius, 
$s_{\perp}$, and the angle between the source trajectory and the binary axis, 
$\alpha$. For a caustic-crossing event, an additional parameter of the source 
radius normalized by the Einstein radius, $\rho_{\star}$ (normalized source 
radius), is required to account for the finite-source effect \citep{dominik95, 
gaudi99, gaudi02, pejcha09}.

\begin{figure}[th]
\epsscale{0.8}
\plotone{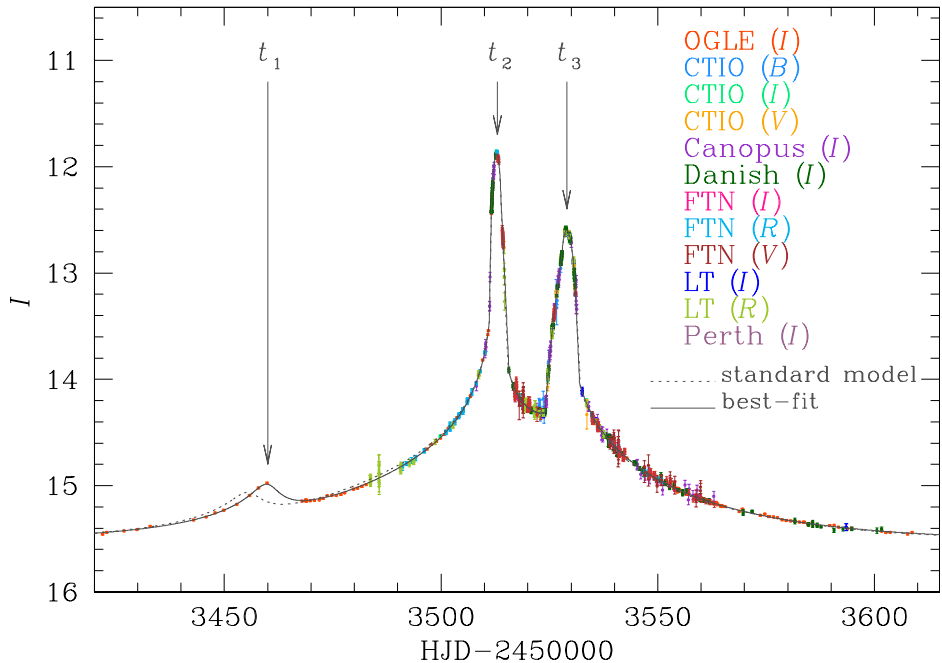}
\caption{\label{fig:one}
Light curve of the microlensing event OGLE-2005-BLG-018. Also presented 
our light curves from modelings based on the standard binary parameters 
(dotted curve) and considering the parallax and orbital effects (solid 
curve).  
}\end{figure}

However, modeling binary-lens light curves with the standard parameters
is occasionally not adequate to precisely describe light curves. 
This is because light curves are subject to various higher-order effects 
that result in deviations from the canonical form. The causes of these 
effects include the motions of the observer, source, and lens during the 
event. The change of the observer's position induced by the orbital motion 
of the Earth around the Sun causes the source motion with respect to the 
lens to deviate from rectilinear, and thus the resulting light curve 
can exhibit long-term deviations. This is known as the ``parallax'' effect
\citep{gould92, refsdal66}. If the source star is a binary, the source 
trajectory can also be affected by the orbital motion of the source over 
the course of the event. This is known as the ``xallarap'' effect 
\citep{han97}, i.e., reverse order of the spell of ``parallax''. 
Finally, the binary nature of the lens implies that the positions of the 
lens components vary in time due to their orbital motion. This ``orbital'' 
effect causes not only the change of the source position with respect to 
the lens but also the magnification pattern because the projected binary 
separation changes in time \citep{dominik98,ioka99,albrow00, rattenbuty02}.  
The deviations in lensing light curves caused by these second-order effects 
are usually very small and thus difficult to measure.  However, when they 
are measured, they provide information that allows to better constrain the 
lens system. For example, if the deviation caused by the parallax effect 
is measured, it is possible to determine the physical parameters of the 
lens mass, distance to it, and the projected separation in physical units 
\citep{gould92}.  If the orbital effect is measured, one can further 
constrain the lens system by determining the orbital parameters and 
intrinsic separation between the lens components \citep{gaudi08, dong09, 
bennett10a, penny10, yee10, skowron10}.

In this paper, we analyze the light curve of the binary-lensing event
OGLE-2005-BLG-018 by combining all available data obtained from survey
and follow-up observations. The event was analyzed before by \citet{skowron07} 
based on the OGLE data and their model shows noticeable deviations. Our 
preliminary modeling also indicated that a model based on the standard binary 
parameters is not adequate enough to precisely describe the observed light 
curve, suggesting the need to consider higher-order effects. We conduct 
analysis of the light curve considering various second-order effects and 
present the constraints of the lens system obtained from modeling.

\section{Observation}

The event OGLE-2005-BLG-018 occurred on a Galactic bulge star located at 
$(\alpha, \delta)_{\rm J2000}=(17^{\rm h}51^{\rm m} 23^{\rm s} \hskip-2pt .53, 
-29^\circ 39' 22''\hskip-2pt .8)$, which corresponds to the Galactic coordinates 
$(l,b)=(0.03^\circ, -1.45^\circ)$.  The event was detected by the Optical 
Gravitational Lensing Experiment \citep{udalski03} using the 1.3 m Warsaw 
telescope of Las Campanas Observatory in Chile.  An anomaly alert was issued 
on 2005 Mar 31 by the OGLE group.  In addition, a series of real-time models 
were issued by M.\ Dominik (2010, private communication).  Following the alert 
and models, the event was intensively observed by follow-up groups including 
the Probing Lensing Anomalies Network \citep{beaulieu06}, RoboNet \citep{tsapras09}, 
and Microlensing Follow-Up Network \citep{gould06} by using six telescopes 
located on four different continents. The telescopes used for follow-up 
observations include 2.0 m Faulkes Telescope N. (FTN) in Hawaii, 2.0 m Liverpool 
Telescope (LT) in La Palma, Spain, 1.0 m of Mt.\ Canopus Observatory in Australia, 
and 1.54 m Danish Telescope of La Silla Observatory in Chile, 0.6 m of Perth 
Observatory in Australia, and 1.3 m SMARTS telescope of CTIO in Chile.  Thanks 
to the follow-up observations, the light curve was very densely resolved.

Figure \ref{fig:one} shows the light curve of the event. It is characterized 
by three distinctive features, occurring at ${\rm HJD} \sim 2453460\ (t_1)$, 
$2453512\ (t_2)$, and $2453528\ (t_3)$.  The two adjacent peaks at $t_2$ 
and $t_3$ are strong while the other peak at $t_1$ is relatively weak and 
separated from the strong features.

\section{Modeling of Second-Order Effects}

We first conduct modeling of the light curve with the set of standard binary 
parameters.  As shown in Figure \ref{fig:one}, the best-fit light curve from 
this initial modeling shows noticeable deviations from the observed light 
curve especially near the part of the light curve around the weak feature 
although the model light curve describes the two strong features relatively 
well.  Investigation of the lens system geometry obtained from the standard 
modeling indicates that the projected separation between the binary lens 
components is smaller than the Einstein radius, i.e., $s_\perp<1$. In this 
case, the resulting caustics are composed of 3 segments, where one large 
central caustics is located around the center of mass of the binary and the 
other two small caustics are located away from the central caustic 
\citep{schneider86}.\footnote{In more precise term, the number of caustic
is 3 when $s_\perp < (1+q)^{1/4}(1+q^{1/3})^{-3/4}$ and the three caustics 
merge into a single one as the separation becomes equivalent to the Einstein 
radius.} The model also indicates that the two strong features 
of the light curve were produced by two successive crossings of the source 
over the central caustic and the weak feature was produced by the approach 
to one of the two peripheral caustics. Events produced by such a lens system 
are susceptible to the orbital effect because the peripheral caustic moves 
considerably even for a small shift of the binary axis. In addition, the 
long duration of the event, which lasted $\sim 200$ days, raises the need 
to consider both the parallax and xallarap effects. We, therefore, conduct 
modelings of the light curve considering these second-order effects as well.

Considering the parallax effect into modeling requires to include two 
additional parameters of $\pi_{{\rm E},N}$ and $\pi_{{\rm E},E}$. These 
parameters represent the two components of the lens parallax vector 
$\pivec_{\rm E}$ projected on the sky in the north and east equatorial 
coordinates, respectively. The direction of the parallax vector corresponds 
to that of the lens-source relative motion in the frame of the Earth at a 
specific time of the event.  We set the reference time at the moment of 
the second perturbation peak, i.e., $t_2$. The size of the parallax vector 
corresponds to the ratio of the Earth's orbit to the Einstein radius 
projected on the observer observer's plane, i.e., $\pi_{\rm E}={\rm AU}/
[r_{\rm E} D_{\rm S}/(D_{\rm S}-D_{\rm L})]$.

Under the assumption of a circular orbit and a very faint binary companion, 
the xallarap effect is described by 5 parameters. They are the orbital period 
of the source, $P_{\rm S}$, inclination, $i_{\rm S}$, phase angle, $\psi$, 
and the two components of the xallarap vector in the north and east direction, 
$\xi_{{\rm E},N}$ and $\xi_{{\rm E},E}$.  The magnitude of the xallarap vector
$\xivec_{\rm E}$ corresponds to the ratio of the source star's orbit to 
the Einstein radius projected on the source plane.

\begin{figure}[th]
\epsscale{0.8}
\plotone{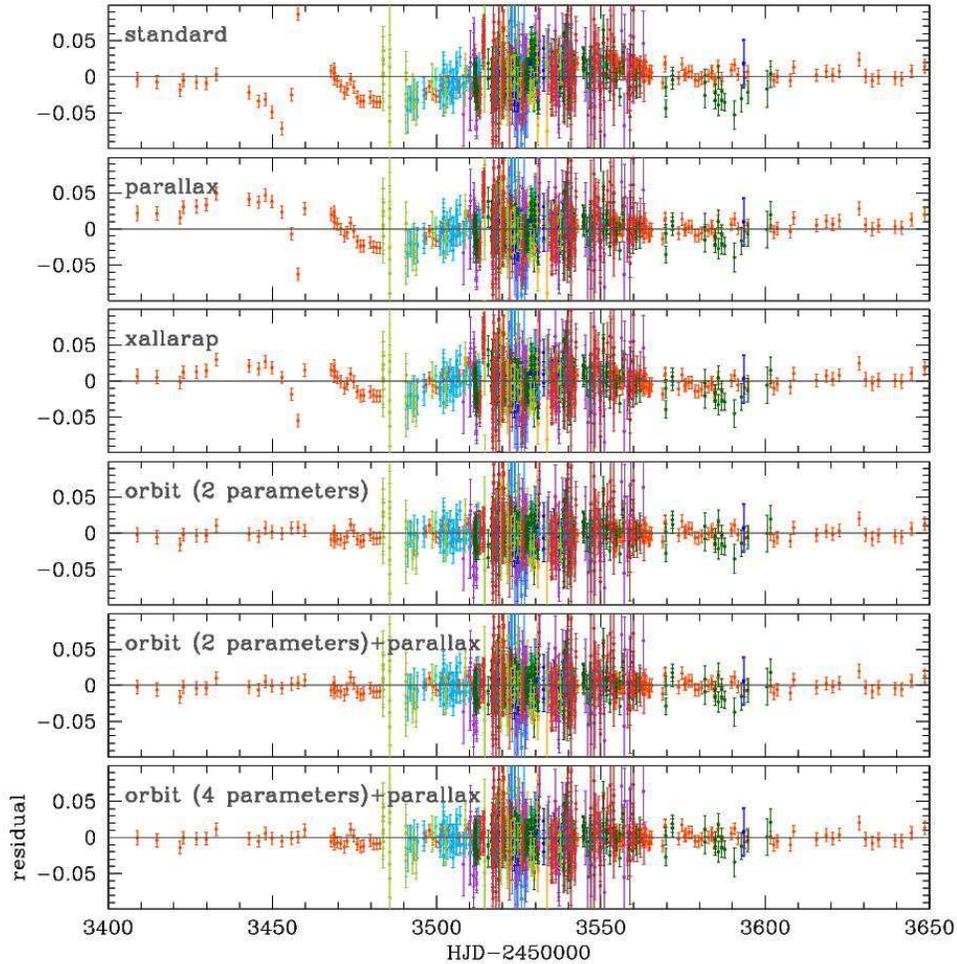}
\caption{\label{fig:two}
Residual of data from 6 different models.
}\end{figure}

To account for the orbital effect, we consider 2 types of parameterization. 
The first one is based on the approximation that the binary lens rotates 
with a constant angular speed and the projected separation between its 
components changes with a constant rate. This parameterization requires 
two parameters of $d\alpha/dt$ and $ds_\perp/dt$, which represent the 
changing rates of the angle between the binary axis and the source 
trajectory and the projected separation between the lens components, 
respectively. In the second parameterization, we fully consider the 
Keplerian orbital motion. This requires to include 2 extra parameters 
in addition to the orbital parameters used in the first type of 
parameterization. These additional parameters are $s_\parallel$ and 
$ds_\parallel/dt$, where $s_\parallel$ represents the line-of-sight 
separation between the binary components in units of $\theta_{\rm E}$ 
and $ds_\parallel/dt$ represents its rate of change.  For the full 
description of the orbital lensing parameters, see the summary in the 
Appendix of \citet{skowron10}.

With these parameterizations, we test 6 different models.  The first 
model is based on standard binary parameters with a static lens and 
source and no parallax motion of the Earth (standard model). The second 
and third models include the parallax and xallarap effects, respectively. 
The fourth model includes the orbital effect of the lens with 2 parameters 
of $d\alpha/dt$ and $ds_\perp/dt$.  In the last two models, we consider 
both the parallax and orbital effects where the orbital effect is described 
by 2 and 4 parameters, respectively.  We note that 4-parameter orbital 
modeling must include the parallax parameters whereas 2-parameter orbital 
modeling does not necessarily needs this.

We search for the solution of the best-fit parameters of the individual models 
by minimizing $\chi^2$ in the parameter space. For binary-lensing modeling, 
this is a very complex procedure due to several reasons. First, the complexity 
of the $\chi^2$ surface in the parameter space makes it difficult to rule out 
the possibility of the existence of local minima \citep{dominik99}, implying 
that even if a plausible model is found, it is difficult to be sure the solution 
is the correct one. This makes it difficult to use a simple downhill approach to 
search for solutions. Second, modeling is further complicated by the sheer size 
of the parameter space. The large number of parameters implies that brute-force 
searches for solutions are very difficult and extremely time-consuming.  To 
resolve the degeneracy problem but avoiding searches throughout all parameter 
space, we use a hybrid approach in which grid searches are conducted over the 
space of a set of parameters and the remaining parameters are searched by using 
a downhill approach.  We choose $s_\perp$, $q$, and $\alpha$ as the grid parameters 
because they are related to the light curve features in a complex way such that 
a small change in the values of the parameters can lead to dramatic changes in 
the resulting light curve. On the other hand, the other parameters are more 
directly related to the light curve features and thus they can be searched for 
by using a downhill approach. For the $\chi^2$ minimization in the downhill 
approach, we use the Markov Chain Monte Carlo (MCMC) method.

Another difficulty in binary-lensing modeling arises due to large computation. 
Most binary-lens events exhibit perturbations induced by caustic crossings or 
approaches during which the finite-source effect is important. Calculating finite 
magnifications involves a numerical method, which requires heavy computations. 
Considering that modeling requires to produce a large number of light curves of 
trial models, it is important to apply an efficient method for magnification 
calculations. We accelerate the computation first by minimizing computations in 
the numerical method and second by restricting the numerical computation only when 
it is necessary. The numerical method applied for finite magnification computations 
is based on the ray-shooting method. In this method, a large number of rays are 
uniformly shot from the image plane, bent according to the lens equation, land on 
the source plane, and then the magnification is computed as the ratio of the number 
density of rays on the source surface to the density on the image plane. In this 
process, we reduce the number of rays required for magnification computations by 
shooting only ones arriving at the region around the caustics.  We further restrict 
numerical computations by applying a simple analytic hexadecapole approximation for 
finite magnifications \citep{pejcha09, gould08} unless the source is located 
very close to the caustics.

We incorporate the effect of the limb-darkening of the source star
surface when we compute the finite-source magnification. The surface
brightness is modeled by
\begin{equation}
S_{\lambda}={F_{\lambda} \over \pi \theta_\star^2}
\left[ 1-\Gamma_\lambda\left( 1-{3 \over2} \cos \theta \right)\right],
\label{eq3}
\end{equation}
where $\Gamma_\lambda$ is the linear limb-darkening coefficients,
$F_\lambda$ is the flux, and $\theta$ is the angle between the normal to the
source star's surface and the line of sight toward the star. From the
color of the source star measured from the location on the
color-magnitude diagram, it is found that the source is a clump giant.
We, therefore, use the limb-darkening coefficients of $\Gamma_V=0.708$,
$\Gamma_I=0.613$, and $\Gamma_R=0.508$ by adopting the values from
\citet{claret00} under the assumption that 
$v_{\rm turb}=2\ {\rm km}\ {\rm s}^{-1}$, $\log (g/g_\odot)=-1.9$, and 
$T_{\rm eff}=4750$ K.

\begin{figure}[th]
\epsscale{0.4}
\plotone{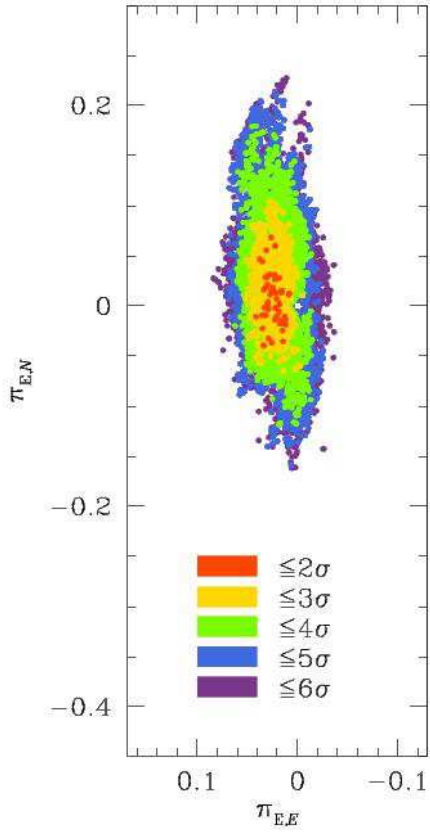}
\caption{\label{fig:three}
Scatter plot of Markov chains in the space of the parallax parameters 
$\pi_{{\rm E},E}$ and $\pi_{{\rm E},N}$ for the best-fit model.
Different colors of points represent $\Delta \chi^{2}$ from the minimum.
}\end{figure}

\begin{figure}[th]
\epsscale{0.8}
\plotone{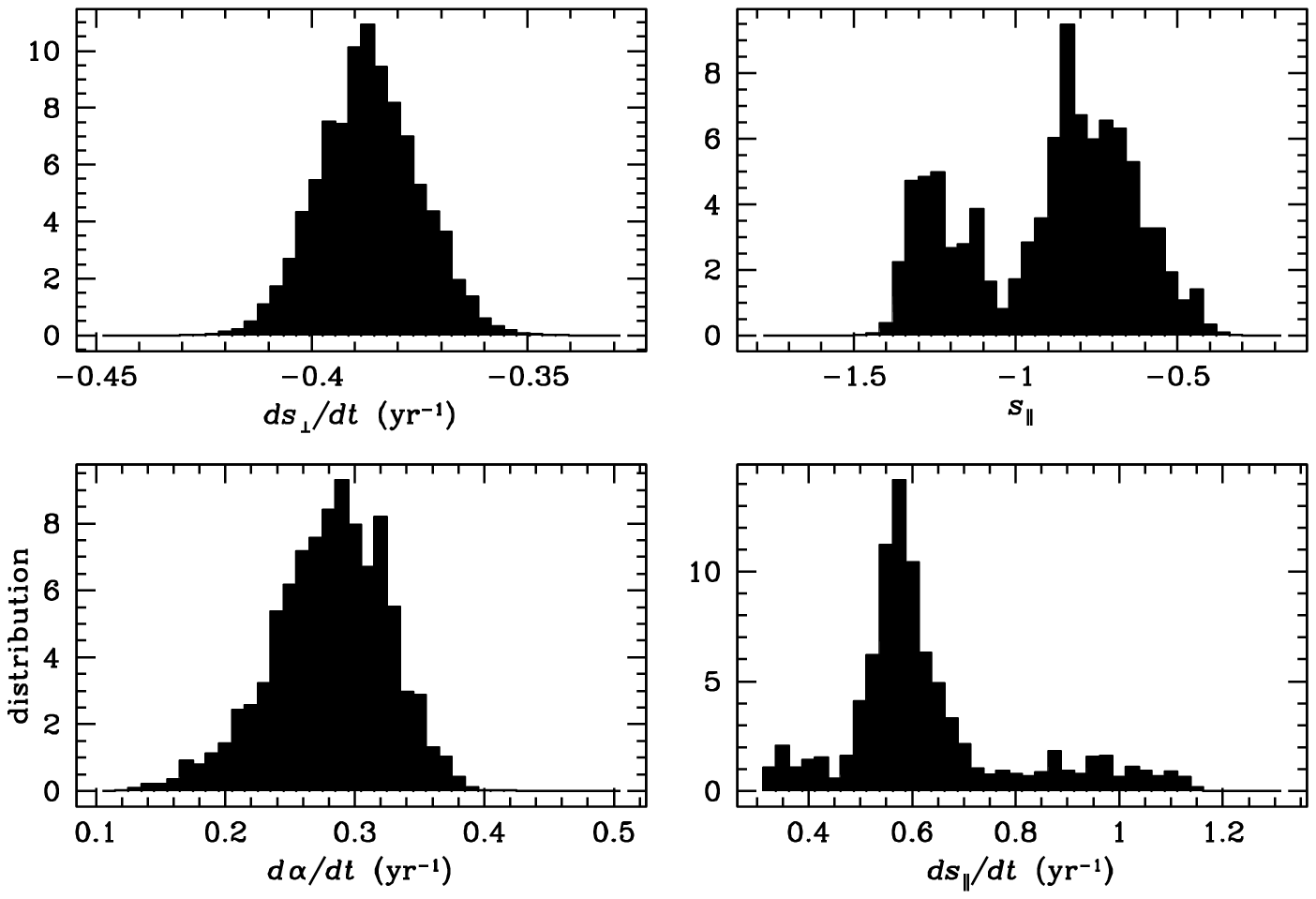}
\caption{\label{fig:four}
Histograms of lens orbital parameters obtained from the 4-parameter orbital fit.
}\end{figure}

\section{Result}

In Table \ref{table:one}, we present the solutions found for the 6 tested 
models.  In Figure \ref{fig:two}, we also present the residuals of the data 
from the best-fit light curves of the individual models. We find that neither 
the parallax nor the xallarap effect alone is enough to precisely describe 
the observed light curve although the models considering these effects improve 
the fit from the standard model with $\Delta\chi^2/{\rm dof}\sim 1113/1600$ 
and 1151/1597, respectively.  For both models, the fits are still poor near 
the weak feature of the light curve.

On the other hand, it is found that the light curve is well described by the 
model including the orbital effect.  We find that the fit with 2 orbital 
parameters is better than the standard model by $\Delta\chi^2/{\rm dof}\sim 
1853/1600$. It is also better than the best-fit parallax and xallarap models 
by $\Delta\chi^2/{\rm dof}\sim 740/1600$ and 702, respectively.  Therefore, 
we find that the dominant second-order effect for the deviation between the 
observed data and standard model is the orbital motion of the lens. As mentioned 
in the previous section, the importance of the orbital effect was expected due 
to the specific geometry of the lens system in which the source trajectory passes 
the central and the peripheral caustics of a close binary.  In this sense, the 
event  has much in common with the event MACHO 97-BLG-41 for which the orbital 
effect was measured for the first time \citep{albrow00}.

The Einstein radius is measured from the normalized source radius, 
$\rho_{\star}$, determined from modeling combined with the angular radius of the 
source star, $\theta_{\star}$, i.e., $\theta_{\rm E}=\theta_{\star}/\rho_{\star}$. 
We measure the angular source radius by first measuring the de-reddened color of 
the source star and converting it into radius by using the relation between the 
color and angular radius of \citet{kervella04}. For the calibration of the magnitude 
and color of the source star, we use the centroid of bulge clump giants in the 
color-magnitude diagram as a reference under the assumption that the source and 
clump giants experience the same amount of extinction \citep{yoo04}. The measured 
Einstein radius is $\theta_{\rm E}=0.506\pm0.044$ mas.

\begin{figure}[th]
\epsscale{0.6}
\plotone{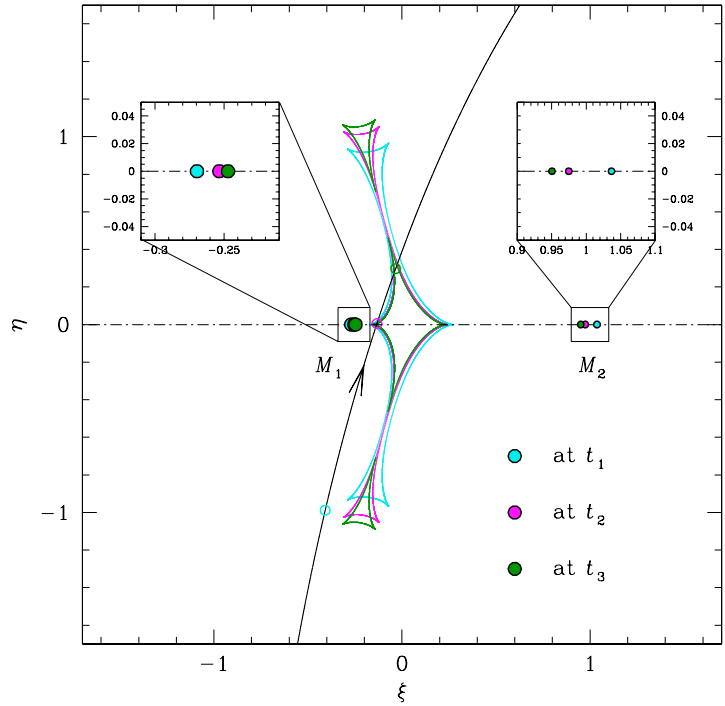}
\caption{\label{fig:five}
Geometry of the binary-lens system responsible for the lensing event OGLE-2005-BLG-018.
The filled dots represent the locations of the lens components at the times of 3 
different major perturbations, where the bigger dots represent the heavier lens 
component. The insets show the zoom of the lens positions. The closed figure 
composed of concave curves represent the caustics where the colors correspond 
to those of the lens. The line with an arrow represents the source trajectory. 
The coordinates ($\xi$,$\eta$) are centered at the center of mass of the binary 
and all lengths are scaled by the Einstein radius corresponding to the total mass 
of the binary lens.
}\end{figure}

Although the importance of the orbital motion of the lens in describing the 
observed light curve is identified, we conduct additional modeling considering 
both orbital and parallax effects in order to see the possibility of further 
improvement of the fit and to constrain physical parameters of the lens system.
The orbital effect is considered by both models with 2 and 4 parameters.
From the model with the parallax plus 2 orbital parameters, it is found that 
adding the parallax effect does not improve the fit significantly.  This could 
be because the parallax is poorly constrained or because it constrained close 
to zero.  From the other modeling with 4 orbital parameters, it is found that
the latter holds.  In Figure \ref{fig:three} and \ref{fig:four}, we present 
the scatter plot of Markov chains in the $\pi_{{\rm E},E}$-$\pi_{{\rm E},N}$ 
parameter space and the histograms of the microlens orbital parameters, respectively.
They show that the parallax and orbital parameters are reasonably well constrained. 
We measure the lens parallax of $\pi_{\rm E}=0.028\pm0.010$.  A small parallax 
value suggests that either the lens is heavy or it is located away from the Earth. 
On the top of the light curve in Figure \ref{fig:one}, we present the best-fit 
light curve for this model.  In Figure \ref{fig:five}, we also present the source 
trajectory with respect to the caustic.  We note that the caustic shape varies 
with time. We present three sets of caustics corresponding to the times of $t_1$, 
$t_2$, and $t_3$. Also marked are the positions of the lens components at the 
corresponding times.

We determine the physical and orbital parameters of the lens system based 
on the measured lensing parameters.  This requires to adopt a value of the 
Einstein radius in the modeling process including 4 orbital parameters.
We adopt this value as the one measured from the model with 2 orbital 
parameters, i.e., $\theta_{\rm E}=0.506\pm0.044$ mas.  In principle, the 
value of $\theta_{\rm E}$ could change as more parameters are added.  However, 
this change is usually very small because the constraint of $\rho_\star$, from 
which $\theta_{\rm E}$ is measured, is provided by the very localized region 
of the light curve where the finite-source effect is important, while the 
orbital and parallax effects are constrained from the overall shape of the 
light curve.  As a result, the physical and orbital parameters are barely 
affected by the adopted value of $\theta_{\rm E}$.  With the Einstein radius 
and the lens parallax determined from modeling, the mass and distance to the 
lens are determined by
\begin{equation}
M=M_1 + M_2 = {\theta_{\rm E} \over \kappa\pi_{\rm E}}
\label{eq4}
\end{equation}
and
\begin{equation}
D_{\rm L}={{\rm AU} \over \pi_{\rm E}\theta_{\rm E}+\pi_{\rm S}}, 
\label{eq5}
\end{equation}
respectively, where $M_1$ and $M_2$ are the masses of the heavy and light 
components of the lens, respectively, and $\pi_{\rm S}={\rm AU}/D_{\rm S}$ 
is the parallax of the source star.

In addition to the constraints provided by the light curve itself, the lens 
system can also be constrained by the blended flux.  This is because the 
flux from the lens cannot exceed the measured blended flux.  We find that 
the blended flux is negligible compared to the flux from the source star.
Even considering that the source is a giant, this provides the constraint 
that the primary of the lens should be a main-sequence star.  
Therefore, we set the upper mass limit of the primary as $\sim 1.3\ M_\odot$, 
and thus the total mass of the lens should be $\leq 2.0\ M_\odot$.

\begin{figure}[th]
\epsscale{0.8}
\plotone{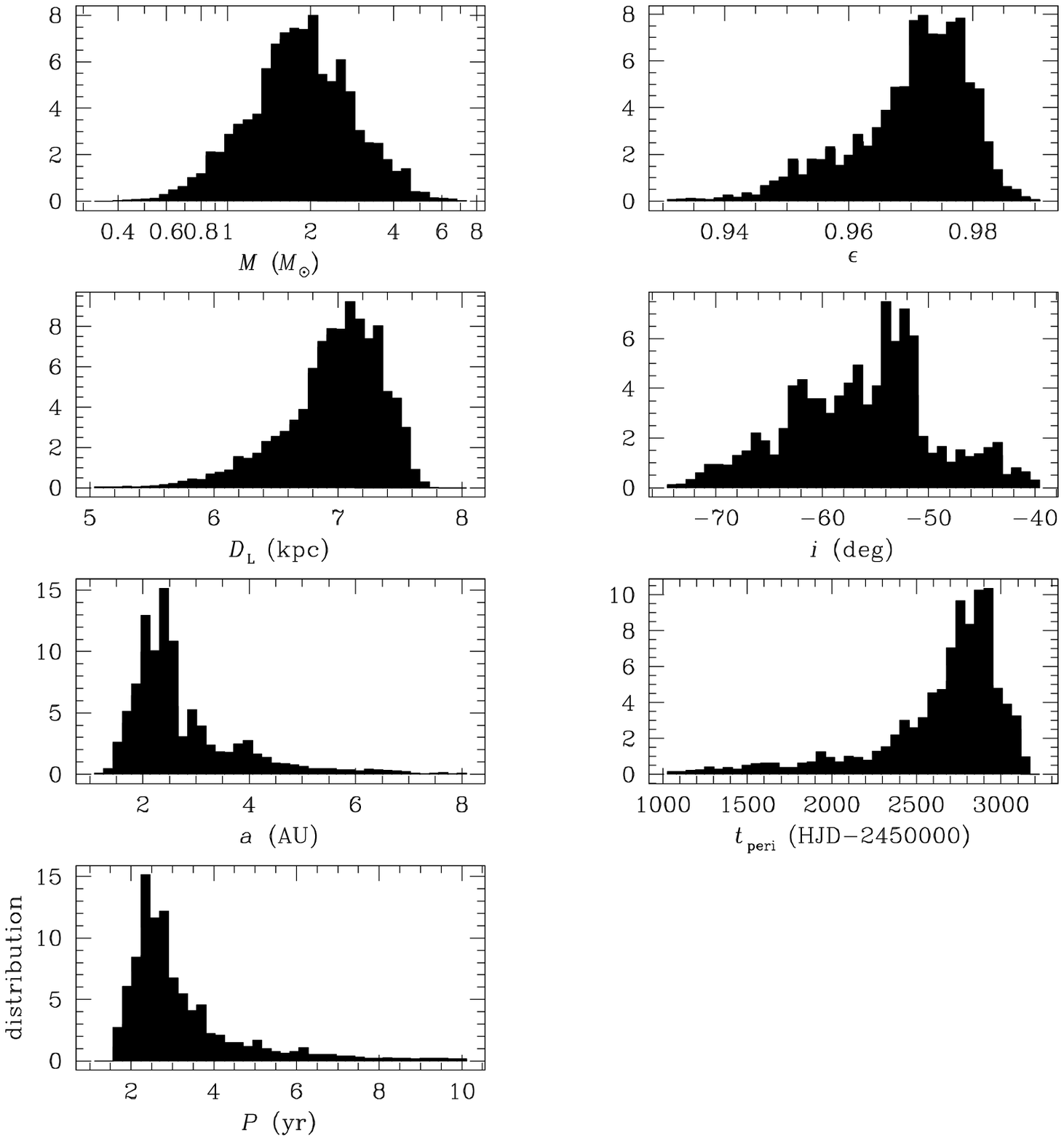}
\caption{\label{fig:six}
Histograms of the physical and orbital parameters, The light and dark 
shaded histograms are with and without the constraint from blended flux, respectively.
}\end{figure}

In Figure \ref{fig:six}, we present the distributions of the physical 
and orbital parameters determined from modeling. The histograms are 
based on the chains obtained from MCMC running, where the dark and light 
shaded ones are with and without the constraint from blended flux, 
i.e.\ $M\leq 2.0 M_\odot$, respectively. We measure the physical and 
orbital parameters and their uncertainties as the mean and standard deviation 
of the values in the chains and list them in Table \ref{table:two}.  It is 
found that OGLE-2005-BLG-018 was produced by a binary lens located in the 
Galactic bulge with a distance to the lens of $D_{\rm L}=6.7\pm 0.3$ kpc.  
The lens is composed of two main-sequence stars with masses 
$M_1=0.9\pm 0.3\ M_\odot$ and $M_2=0.5\pm 0.1\ M_\odot$.  The mass of the 
lens system is consistent with the restriction of $M=M_1+M_2< 2\ M_\odot$, 
that was given by the blended flux.  The two lens components are separated 
by a semi-major axis of $a=2.5\pm 1.0\ {\rm AU}$ and orbiting each other 
with an orbital period of $P=3.1\pm 1.3$ yr.

\section{Conclusion}

We analyzed the light curve of OGLE-2005-BLG-018 based on the combined
data from survey and follow-up observations. The light curve shows
noticeable deviations from the best-fit model based on the standard
binary parameters. From modeling including various higher-order effects,
we found that the dominant second-order effect for the deviation is the
orbital motion of the lens. Based on modeling considering full-Keplerian
orbital motion and the parallax effect, we were able to measure the
physical and orbital parameters of the lens system. Detections of
higher-order effects and determinations of the physical lens parameters 
were possible due to the well-resolved light curve covering all three major
perturbations. Unfortunately, events with such detections of higher-order 
effects are rare for events detected in current lensing experiments based 
on the survey/follow-up mode.  This is because it is difficult to resolve 
perturbations from survey observations alone and even if perturbations are 
detected and an alert is issued for follow-up observations, it is unavoidable 
to miss part of the perturbation due to the time gap between the alert and 
the initiation of follow-up observations.

However, the advent of next-generation experiments based on ground and in 
space will make it possible to routinely measure higher-order effects for 
a large fraction of lensing events. The OGLE and MOA survey groups have 
recently upgraded their cameras with wider field of view.  The Korea 
Microlensing Telescope Network (KMTNet) is a funded project that plans to
achieve $\sim 10$ minute sampling of all lensing events using a network 
of 1.6 m telescopes to be located in three different continents of southern 
hemisphere with cameras of 4 ${\rm deg}^2$ field of view. Furthermore, there 
are planned lensing surveys in space including EUCLID \citep{beaulieu10} and 
WFIRST \citep{bennett10b}, that are proposed to ESA and recommended as the 
top ranked large space missions of NASA for the next decade, respectively. 
With these experiments come on line, nearly all events will be densely 
observed, making it possible to routinely measure the higher-order effects 
and thus to constrain the physical parameters of lenses.

\acknowledgments 
This work was supported by Creative Research Initiative program
(2009-0081561) of National research Foundation of Korea.  
Work by A.G. was supported by NSF grant AST-0757888.
The OGLE project has received funding from the European Research Council
under the European Community's Seventh Framework Programme
(FP7/2007-2013) / ERC grant agreement no.\ 246678 to AU.

\headsep=70pt
\begin{sidewaystable}[ht]
\caption{Best-fit Parameters for 6 Tested Models\label{table:one}}
\begin{tabular}{lcccccc}
\hline\hline 
\multicolumn{1}{c}{parameters} &
\multicolumn{6}{c}{model} \\
\multicolumn{1}{c}{} &
\multicolumn{1}{c}{standard} &
\multicolumn{1}{c}{parallax} &
\multicolumn{1}{c}{xallarap} &
\multicolumn{1}{c}{orbit} &
\multicolumn{1}{c}{parallax +} &
\multicolumn{1}{c}{parallax +} \\
\multicolumn{1}{c}{} &
\multicolumn{1}{c}{} &
\multicolumn{1}{c}{} &
\multicolumn{1}{c}{} &
\multicolumn{1}{c}{(2-parameters)} &
\multicolumn{1}{c}{orbit (2 parameters)} &
\multicolumn{1}{c}{orbit (4 parameters)} \\
\hline
$\chi^2/{\rm dof}$                     & 3465.1/(1602)      & 2352.2/(1600)      & 2314.0/(1597)      & 1611.8/(1600)      & 1610.506/(1598)     & 1607.121/(1596)    \\
$t_0$ (HJD-2450000)                    & 3514.565$\pm$0.007 & 3514.577$\pm$0.008 & 3514.546$\pm$0.011 & 3514.931$\pm$0.014 & 3514.906$\pm$0.0143 & 3514.927$\pm$0.016 \\
$u_0$                                  & 0.124$\pm$0.001    & 0.122$\pm$0.001    & 0.126$\pm$0.001    & 0.127$\pm$0.001    & 0.127$\pm$0.001     & 0.128$\pm$0.001    \\
$t_{\rm E}$ (days)                     & 50.44$\pm$0.09     & 51.629$\pm$0.072   & 49.921$\pm$0.173   & 52.297$\pm$0.083   & 52.337$\pm$0.115    & 52.130$\pm$0.159   \\
$s_{\perp}$                            & 0.715$\pm$0.001    & 0.702$\pm$0.001    & 0.705$\pm$0.001    & 0.722$\pm$0.001    & 0.722$\pm$0.001     & 0.724$\pm$0.001    \\
$q$                                    & 0.521$\pm$0.001    & 0.528$\pm$0.002    & 0.555$\pm$0.004    & 0.539$\pm$0.003    & 0.536$\pm$0.003     & 0.539$\pm$0.002    \\
$\alpha$ (rad)                         & 4.998$\pm$0.001    & 5.002$\pm$0.001    & 4.998$\pm$0.002    & 5.028$\pm$0.001    & 5.025$\pm$0.002     & 5.026$\pm$0.002    \\
$\rho_\star$                           & 0.025$\pm$0.001    & 0.025$\pm$0.001    & 0.025$\pm$0.001    & 0.025$\pm$0.001    & 0.025$\pm$0.001     & 0.026$\pm$0.001    \\
$\pi_{{\rm E},N}$                      & --                 & 0.115$\pm$0.011    & --                 & --                 &-0.044$\pm$0.030     &-0.011$\pm$0.028    \\
$\pi_{{\rm E},E}$                      & --                 & 0.342$\pm$0.008    & --                 & --                 &-0.006$\pm$0.010     & 0.021$\pm$0.010    \\
$\xi_{{\rm E},N}$                      & --                 & --                 &-0.039$\pm$0.004    & --                 & --                  & --                 \\
$\xi_{{\rm E},E}$                      & --                 & --                 &-0.039$\pm$0.001    & --                 & --                  & --                 \\
$\psi$                                 & --                 & --                 & 3.98$\pm$0.25      & --                 & --                  & --                 \\ 
$i_{\rm S}$                            & --                 & --                 & 1.50$\pm$0.10      & --                 & --                  & --                 \\
$P_{\rm S}$ (yr)                       & --                 & --                 & 0.45$\pm$0.01      & --                 & --                  & --                 \\
$d{s_{\perp}}/dt$ ($\rm yr^{-1}$)      & --                 & --                 & --                 &-0.409$\pm$0.009    &-0.387$\pm$0.011     &-0.389$\pm$0.013    \\
$d{\alpha}/dt$ ($\rm yr^{-1}$)         & --                 & --                 & --                 & 0.272$\pm$0.022    & 0.328$\pm$0.049     & 0.315$\pm$0.046    \\
$s_{\parallel}$                        & --                 & --                 & --                 & --                 & --                  &-0.832$\pm$0.180    \\
$d{s_{\parallel}}/dt$ ($\rm yr^{-1}$)  & --                 & --                 & --                 & --                 & --                  & 0.581$\pm$0.161    \\
\hline
\end{tabular}
\\
\\Notes. HJD'=HJD-2450000.
\end{sidewaystable}

\begin{deluxetable}{lc}
\tablecaption{Physical and Orbital Parameters\label{table:two}}
\tablewidth{0pt}
\tablehead{
\multicolumn{1}{c}{parameter} &
\multicolumn{1}{c}{values} \\
}
\startdata
$M_{\rm total}$ ($M_{\odot}$)    & 1.38$\pm$0.39  \\ 
$M_1$ ($M_{\odot}$)              & 0.90$\pm$0.25  \\
$M_2$ ($M_{\odot}$)              & 0.48$\pm$0.14  \\
$D_{\rm L}$ (kpc)                & 6.74$\pm$0.32  \\ 
$a$ (AU)                         & 2.46$\pm$0.97  \\ 
$P$ (yr)                         & 3.10$\pm$1.30  \\ 
$\epsilon$                       & 0.97$\pm$0.01  \\     
$i$ (deg)                        &-55.01$\pm$6.69 \\ 
$t_{\rm peri}$ (HJD')            & 2670$\pm$352   \\ 
\enddata
\end{deluxetable}

\end{document}